\documentclass[sigconf,authorversion,nonacm,review=false,timestamp=false]{acmart}

\AtBeginDocument{%
  }

\usepackage{listings}
\usepackage{listings-golang}

\definecolor{base}{gray}{0} 
\definecolor{comment}{gray}{0.2} 
\definecolor{string}{rgb}{0.5,0.1,0.20} 

\setcopyright{acmlicensed}
\copyrightyear{2018}
\acmYear{2018}
\acmDOI{XXXXXXX.XXXXXXX}

\usepackage{algorithm}
\usepackage{algpseudocode}

\acmConference[Conference acronym 'XX]{Make sure to enter the correct
  conference title from your rights confirmation emai}{June 03--05,
  2018}{Woodstock, NY}
\acmISBN{978-1-4503-XXXX-X/18/06}

\newcommand{\ctext}[1]{\raise0.2ex\hbox{\textcircled{\scriptsize{#1}}}}

\algnewcommand{\OR}{\algorithmicor}
\algnewcommand{\AND}{\algorithmicand}

\begin{document}

\title{Distributed Tracing for Cascading Changes of Objects in the Kubernetes Control Plane}

\author{Tomoyuki Ehira}
\email{ehira@inet.media.kyoto-u.ac.jp}
\affiliation{%
  \institution{Kyoto University}
  \city{Sakyo}
  \state{Kyoto}
  \country{Japan}
}

\author{Daisuke Kotani}
\email{kotani@media.kyoto-u.ac.jp}
\orcid{0000-0003-4305-8379}
\affiliation{%
  \institution{Kyoto University}
  \city{Sakyo}
  \state{Kyoto}
  \country{Japan}
}

\author{Yasuo Okabe}
\email{okabe@media.kyoto-u.ac.jp}
\affiliation{%
  \institution{Kyoto University}
  \city{Sakyo}
  \state{Kyoto}
  \country{Japan}
}
\renewcommand{\shortauthors}{Ehira et al.}

\begin{abstract}
Kubernetes is a container orchestration system that employs a declarative configuration management approach.
In Kubernetes, each desired and actual state is represented by an ``object'', and multiple controllers autonomously monitor related objects and update their objects towards the desired state in the control plane.
Because of this design, changes to one object propagate to other objects in a chain.
The cluster operators need to know the time required for these cascading changes to complete, as it directly affects the quality of service of applications running on the cluster.
However, there is no practical way to observe this kind of cascading change, including breakdown of the time taken by each change.
Distributed tracing techniques are commonly used in the microservices architecture to monitor application performance, but they are not directly applicable to the control plane of Kubernetes; the microservices architecture relies on explicitly calling APIs on other services, but in Kubernetes the controllers just monitor objects to know when to start processing, and never call functions on other controllers directly.
In this paper, we propose a system that automatically traces changes to objects in the control plane.
Our method adds one identifier, a Change Propagation ID (CPID), to the metadata of an object, and the controller that observes an object change propagates its CPID to the objects that the controller is updated.
When multiple changes need to be merged on an object, a new CPID is generated, and the relationship between the original CPID and the new CPID is sent to the external trace server.
We confirmed that change propagation can be visualized and the required time measured. We also showed that this system's overhead is not significant.
\end{abstract}



\maketitle

\section{Introduction}

A declarative configuration management system is a system that defines and inputs a desired state and automatically updates configurations to to maintain the actual state matches with the desired state.
This architecture is widely used in controllers for large scale systems, and Kubernetes~\cite{kubernetes}\cite{Borg} is a de facto container~\cite{7036275} management system that employs this approach.

In Kubernetes, operators configure resources and their objects by defining only the desired state.
An ``object'' in Kubernetes represents the desired and current state of each cluster function, and a resource is a set of objects of the same function.
In the control plane of Kubernetes, many components called controllers controllers observe the current state of the objects that each controller is in charge of, and constantly perform a process (reconciliation loop) to get them into the desired state.
This mechanism allows the controller to detect when the current state changes due to failures, updates by the operators, or other reasons, and the controller tries to maintain the state of the system in the desired state autonomously.
Since some of these controllers control their resource objects by observing the state of objects in other resources, changes occur in a chain among these resource controllers and objects until status of all related objects are updated.

One of the key metrics in a Kubernetes cluster is the time required to update all related objects when the desired state of one object is changed.
This time directly affects the quality of service of applications running on the cluster.
For example, when requests to an application is suddenly increased, the operators will start more container replicas by updating the number of replicas to handle the requests, but the application cannot handle all the requests until all related objects are updated and additional replicas are launched.
In order to identify where bottlenecks are, we need a mechanism to make cascading changes among related resources visible to operators.

However, there are two challenges for observing these cascading changes. 
First, because the updates of each object is processed autonomously in each controller in the control plane, it is not clear to which controller or object a change is propagated next.
The operator does not specify the sequence of updating each object, but rather, once an object is changed, the various objects are updated autonomously to reflect the change of the object.
Therefore, it is impossible to know (or define) when the change process for all related objects has been completed.
In addition, the objects to be affected by a change of a specific object are different depending on the system, such as the use of custom resources and plugins installed by the operators.
Second, it is difficult to know when a change in one object will cause a change in other objects.
Since each controller is not invoked directly from other controllers but autonomously monitors the current object status and processes it to the desired status, the timing to start the update process of the objects depends on each controller; that is, the updates may be processed one by one just after other objects are updated, or several updates may be processed in batch. 

K-Bench~\cite{k-bench} and ClusterLoader~\cite{clusterloader} are tools to measure the time to handle some predefined types of changes, but they do not have a mechanism for tracing cascading changes between objects in general.
Conventional methods that require manual logging by the operators~\cite{ehira2023monitoring} can trace all resources, but they are costly because they require a deep knowledge of the controller design.
The Kubernetes community has discussed extending the control plane's functionality to allow to measure cascading changes~\cite{KEP647}\cite{KEP-PR-2321}. However, these are still in the discussion because of a problem in handling tracing information to be placed on the object.
Distributed tracing techniques~\cite{Dapper-36356}\cite{202574}\cite{10.5555/647883.738238}\cite{1021256} are commonly used in the microservices architecture to monitor application performance, but they are not directly applicable to the control plane of Kubernetes; the microservices architecture is based on RPCs~\cite{rfc5531} and clearly identify the start and the completion of the requests, but in Kubernetes the controllers just monitor objects to know when to start processing, and never call functions on other controllers directly.

We propose a distributed tracing method in the Kubernetes control plane to facilitate observation of change propagation.
We add change tracing identifiers to object metadata called Change Propagation ID (CPID).
CPID is assigned to objects when processing objects in each controller, and CPID in the objects is propagated to other objects that the controller updates. CPIDs are newly assigned when multiple changes are merged on an object, that is, the object is updated according to multiple objects having different CPIDs.
All logs related to the handling of CPIDs are sent to the external trace server, and the trace server analyses the logs to show the current status of the cascading changes.
Our method incorporates the change propagation logic at the time of implementation of each controller, and the cluster operator can understand relationships between resources and controllers' behavior better to trace cascading changes of objects.

We confirmed that our system can easily trace change propagation and that the performance impact of the system on clusters is not significant.

The contributions of this study are as follows.
\begin{itemize}
  \item We summarize the challenges and requirements for tracing cascading changes specific to the Kubernetes control plane.
  \item This is the first distributed tracing system applied to the Kubernetes control plane, by the idea of combining updating CPIDs in the objects and analysing logs of how CPIDs are updated.
  \item We implement the proposed system and show that changes can be traced in the control plane and that the overhead is acceptable.
\end{itemize}

\section{Related Work}\label{sec:background}

\subsection{Basics of The Kubernetes Control Plane}\label{subsec:control-plane}

In Kubernetes, the control plane is the layer that declaratively manages containers and other computational resources to provide various functions. 
An object represents the desired and current state for each function in the control plane, and a resource is a set of objects with the same functionality.
The objects are stored in the database on the control plane, and Any operations to the objects are performed through the API server, such as creation, updates and deletions.
The following is a list of typical resources that also appear in this paper.

\begin{description}
  \item[Pod] is a logical host running multiple containers that share storage and networking and controls the containers.
  \item[ReplicaSet] represents a set of a specified number of Pods of the same configuration and ensures that the specified number of replicas is always available. If the number of Pods in operation is different from the specified number of replicas, a new Pod will be created or deleted.
  \item[Deployment] offers the functionality of gradually updating container images by changing the number of replicas of multiple ReplicaSets with different container images.
  \item[Service] assigns IP addresses and DNS names to a set of Pods that matches a condition defined by the configuration of Service.
  \item[Endpoints] holds information on IP addresses and port numbers of the Pods pointed to by the corresponding Service.
\end{description}

\begin{figure}[htbp]
  \centering
  \includegraphics[keepaspectratio, width=0.8\hsize]{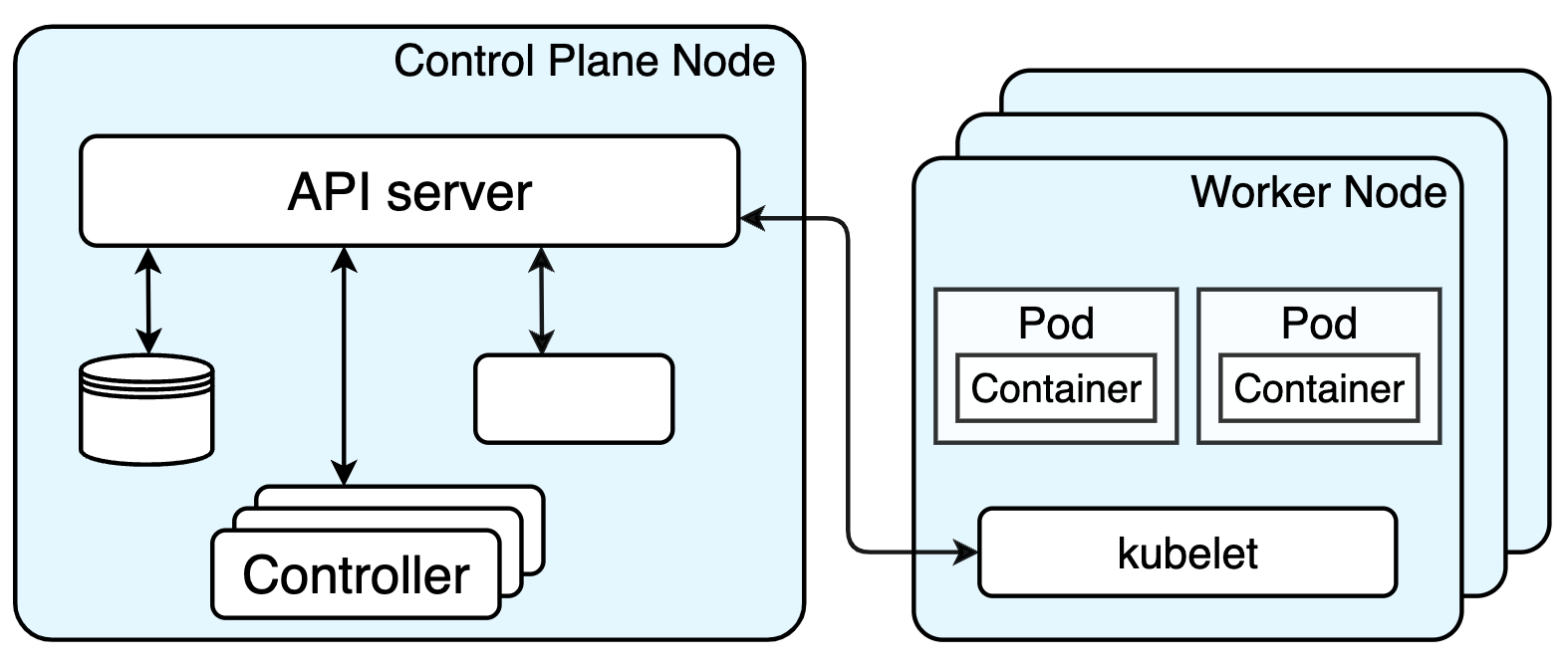}
  \caption{Simplified Kubernetes architecture}
  \label{fig:kubernetes-architecture}
\end{figure}

Figure~\ref{fig:kubernetes-architecture} is a overview of the Kubernetes architecture.
Nodes represent the machines that comprise a Kubernetes cluster, and Pods are assigned to a node and started by a controller called Scheduler.
A kubelet on each node monitors the pod objects deployed on that node, handles the startup and shutdown of containers, and updates the status of objects when container states change.
The API server provides an interface for operations on objects, including notification of an object change to controllers subscribing the monitoring of changes of specific resources.
Each controller manages the objects of the resource it is responsible for.
Each controller also controls the containers or nodes so that the objects are in the desired state when there is a difference between the desired state and the actual state, and this process is called a reconciliation loop.

\begin{figure}[htbp]
  \centering
  \includegraphics[keepaspectratio, width=\hsize]{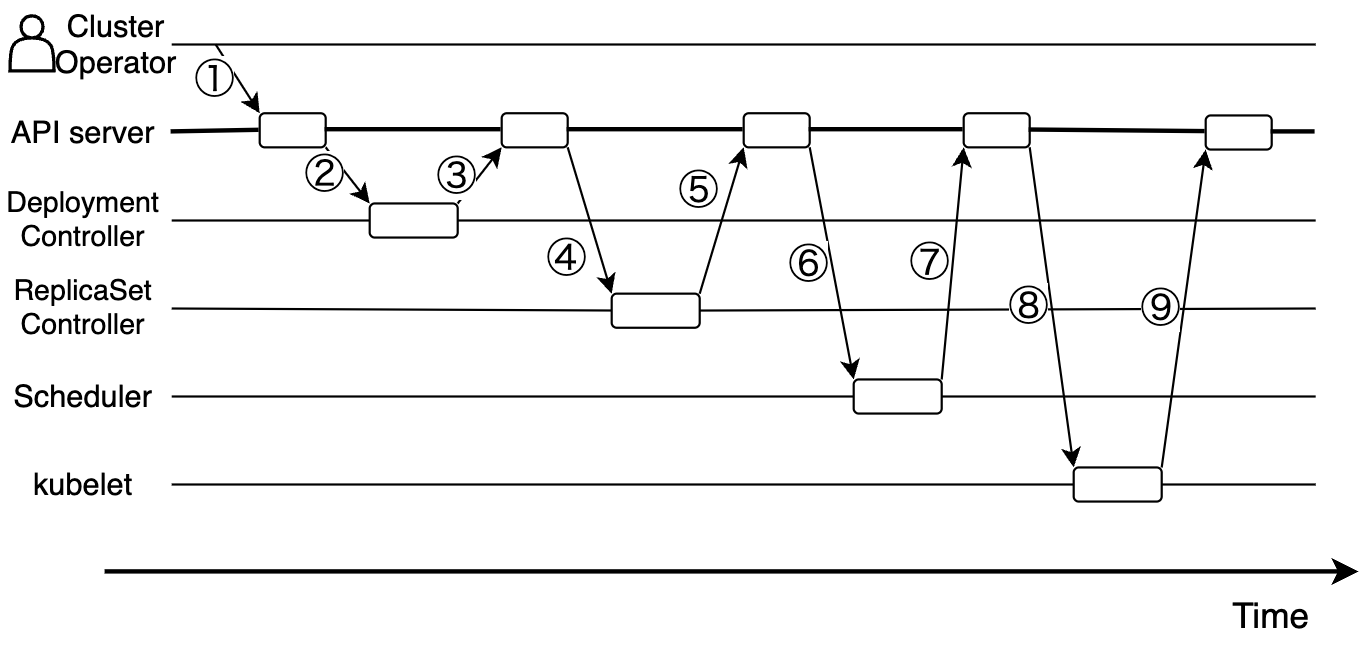}
  \caption{Example of cascading changes: Starting with the creation of the Deployment object, the Deployment controller, ReplicaSet controller, Scheduler, and kubelet observe and process the object changes, and finally the container is created.}
  \label{fig:resource-diagram}
\end{figure}

Figure~\ref{fig:resource-diagram} shows an example of cascading changes across Deployment, ReplicaSet, and Pod. When a Deployment is created, changes are propagated in the control plane until the Pod is started.
\begin{enumerate}
  \item Cluster operator creates a Deployment object.
  \item Deployment controller is notified of the creation of the Deployment object.
  \item Deployment controller creates a ReplicaSet object.
  \item ReplicaSet controller is notified of the creation of the ReplicaSet object.
  \item ReplicaSet controller creates a Pod object.
  \item Scheduler is notified of the creation of the Pod object.
  \item Scheduler determines the node to place the Pod, updates the node name to the Pod object.
  \item kubelet on the selected node is notified of the update of the Pod object.
  \item kubelet starts the container and updates the Pod status.
\end{enumerate}

\subsection{Measuring the Kubernetes Control Plane}\label{subsec:relatedwork-debugging-k8scp}

K-Bench~\cite{k-bench} and ClusterLoader~\cite{clusterloader} allow for measuring the time required for transitions between states of a pod object and the CPU usage during those transitions, mainly for operations on pods from Deployment.
However, these can only be measured between some predefined resources, and it is impossible to monitor and measure the cascading changes in other resources, such as custom resources defined by cluster operators.
Ehira et al.~\cite{ehira2023monitoring} proposes a method to measure the cascading changes among objects that the operator gives relationships in advance. 
Although this method allows to find relationships between objects automatically in part, the operator needs to provide how the objects are related, which requires a significant effort for the operators.

There have been proposals~\cite{KEP647}\cite{KEP-PR-2321} in the Kubernetes community to extend the control plane to automatically trace change propagation by modifying the controllers and other components of the control plane.
These proposes to use a distributed tracing technique similar to this proposal to place trace context on the objects.
These proposals allows to measure the cascading changes without the cluster operator's deep understanding of controllers and a huge amount of additional input to the measurement system.
However, there is a concern that the size of the trace context will continue to grow because there is no method for deleting each trace context.

\subsection{Distributed Tracing}\label{subsec:distributed-tracing}
Distributed tracing~\cite{Dapper-36356}\cite{202574}\cite{10.5555/647883.738238}\cite{1021256} is a technique for tracing how user requests are transmitted between services (applications running in containers) in a distributed system such as a microservice architecture.
User requests are assigned a unique identifier (Trace ID), which is recorded with a timestamp when received by each service, and the same Trace ID is assigned when invoking other services to trace changes.
HTTP headers are used to pass Trace IDs between services, and the W3C Trace Context~\cite{w3c-trace-context} and Zipkin B3 Headers~\cite{zipkin-b3} are widely used for the data format.
This information makes it possible to discover services that are dependencies or bottlenecks.

Information called a span, which includes a trace ID, span ID, and timestamp, is recorded by each service when it receives it, and the context of the change is propagated to other services by adding the trace ID and its span ID when invoking other services.
Tracing requests through distributed tracing requires instrumentation, in which each service assigns a Trace ID to the process when it receives a request and assigns the same Trace ID when it invokes other services.

The motivation of Distributed Tracing is similar to our work, however, due to the difference of architectures between microservices and Kubernetes control plane, we cannot directly apply distributed tracing techniques to the Kubernetes control plane.

Although there is a difference between the propagation of changes in the control plane and the propagation of requests between application services, the purpose of this study is similar in that we want to observe and measure the cascading changes. 
Distributed tracing techniques in microservices assume that the services are commumicated in RPC model, where the services are explicitly called, so the system can explicitly pass Trace IDs and other data along with the content of the requests to the other services.
In Kubernetes control plane, each controller just see the current status of the objects and autonomously handle the change of the objects, so the controllers cannot see who, when, and which parts of the object is updated.

A mechanism to trace the behavior of controllers is especially necessary when extending Kubernetes.
There was a case in which scaling the number of nodes and pods resulted in a significant increase in the time required to create a Pod, as some controllers stopped working due to overloading~\cite{paypal-4k-nodes}.
Also, for example, a third-party extension called Cilium~\cite{cilium} creates an object for every Pod to manage the state of the Pods. In addition, there is a controller that operates based on the status of that object. Thus, the path of change propagation may be prolonged without the cluster operator's awareness.
Thus, especially when expanding the size or functionality of a cluster, abnormal behavior often occurs and the paths of change propagations become complex. Therefore, a distributed tracing mechanism is necessary to trace the behavior of controllers in the control plane to facilitate debugging.

\section{Design of the Proposed Method}\label{sec:proposed-method}

\subsection{Challenges and Requirements}\label{subsec:proposed-method-challenges}
Distributed tracing in Kubernetes control plane has the following challenges and requirements, which is different from the distributed tracing in the traditional RPC model such as in the microservices.

\noindent
\textbf{The completion of the cascading changes is difficult to define.}\quad
In distributed tracing in the typical RPC model, the end of a series of requests can be defined as the point at which the response finally returns to the user.
However, in the control plane, the completion of the cascading changes is hard to define because objects that are affected by a change of the specific object cannot be determined in advance.
Even if we can determine the affected objects completely, it is difficult to determine whether the change made by the controllers is due to the specific change of the specific object because the controllers see the current status of the objects, not how the objects are updated.

For example in Figure~\ref{fig:resource-diagram}, a change in the Deployment object is propagated to the ReplicaSet object and the Pod object. Due to configuration of the control plane, other controllers will see the change of these objects to update the objects that they are in charge of, especially when other plugins are installed in the control plane.

\noindent
\textbf{Multiple changes may be merged while changes are propagated.}\quad
Requests do not join together in distributed tracing in the RPC model, but in control plane change processing, an object can be changed by multiple factors.
Let us think of the case that an object is updated by multiple sources.
The controllers just see the current status of the object, that is, the controllers see the latest status of the object to handle updates.
From the perspective of tracing the cascading changes, handling the objects updated by the multiple sources should be regarded that multiple changes are aggregated and merged into one change.
So, we need to identify which changes are merged into the current object to trace the cascading changes.

For example, when a cluster operator creates a Deployment and before the completion of this creation the automatic scaling function in the control plane changes the number of Pods to be run in the cluster, there should be two changes, and the controllers see the latest number of Pods to be run in the cluster, not the number that the cluster operator first configured.

\noindent
\textbf{Avoid the excessive growth of the trace context}\quad
Each object is persistent in the database until deleted.
Unlike in the typical RPC model, as mentioned above, it is hard to determine when a specific change is completed, so the contexts for tracing the cascading changes may be left on the objects for a long time.
In addition, the size of the contexts will grow without limit if the trace context is added repeatedly due to multiple changes.
Therefore, it is necessary to keep the same size of the trace contexts in the objects regardless of a number of changes or a mechanism that periodically deletes old trace contexts.

In addition, the following should be considered for designing the proposed system.

\noindent
\textbf{Spans and logs related to a given change can be easily retrieved later.}\quad 
This system is intended to investigate the time required for cascading changes. 
Thus, the system needs to collect and present the span of processing and logs of processing due to a certain change in a single step in the Kubernetes control plane, a mechanism in which many components operate autonomously.

\noindent
\textbf{There should be low impact in the presence of non-instrumented controllers.}\quad 
Many third-party controllers are added to the control plane to add functionalities to the control plane.
Because various developers develop these controllers, ensuring that all controllers in the control plane are instrumented is difficult.
Therefore, in an environment where instrumented and un-instrumented controllers coexist, the design must be such that when an object with trace context arrives at an un-instrumented controller, the trace context is preserved and consistent.

\noindent
\textbf{The performance degradation to the cluster is minor.}\quad 
This system must not be extremely slow in propagating changes due to its implementation, such as requiring much more processing time.
It is also essential that the system not consume large amounts of computational resources such as CPU and memory consumption.

\subsection{Main Ideas}\label{subsec:proposed-method-ideas}

This method consists of three main ideas.

\begin{figure}[htbp]
  \centering
  \includegraphics[keepaspectratio, width=0.35\hsize]{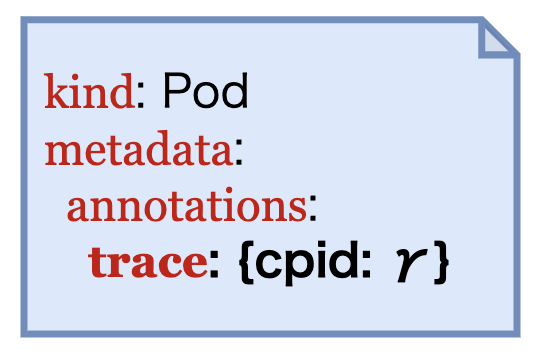}
  \caption{Object with CPID. The trace context is written in the annotation field.}
  \label{fig:object-with-cpid}
\end{figure}

\noindent
\textbf{One CPID per Object}\quad
In the Kubernetes control plane, changes are propagated through object updates, so we put a change tracing identifier called a Change Propagation ID (CPID) on each object.
CPIDs are placed in an annotation field on the metadata, as shown in Figure~\ref{fig:object-with-cpid}.
In our system, CPID is an identifier to trace which object or controller a particular change has reached.
However, unlike trace IDs, a change may be derived from multiple CPIDs due to merging changes.
Here, we assign a new CPIDs when multiple changes are merged, and limit the number of CPIDs to be placed on an object to one.
The details of the merge is discussed next. 

\noindent
\textbf{Merge Process of CPIDs on Change Merging}\quad
When multiple changes are applied to an object, the latest status of the object is regarded as the status that reflects all the changes.
To update other objects that are related to the updated object, controllers see the latest status of the updated object, that is, the controllers refers to the object that all the changes are applied, regardless of the past status of the object.
We call this state as the state that the change is merged, because we can see that all the changes to the object have been applied and the latest status of the object is the results of that.
When a change is merged by updating an object, a new CPID is to be generated.
Precisely speaking, the new CPID is generated and attached to the object when the controller sees different CPIDs between the object that the controller is in charge of, and objects that are managed by other controllers and the controller sees to update the object.

When a new CPID is generated, the list of CPIDs related to this change (including the CPID of the changed object) and the new CPID is sent to the trace server as a mergelog.
Based on the mergelog, the trace server constructs a merge graph with the CPIDs as nodes so that the operators can check which CPIDs were merged into which CPIDs by merging the CPIDs that represent a certain change.
This allows us to continue to trace a change with the context of that change even after that CPID has merged into another CPID.

\noindent
\textbf{Embedding CPIDs in Span and Logs}\quad
As with distributed tracing in regular microservices, each controller in the control plane emit a span logs to tell what kind of processes the controller do and how much time is being spent for a specific processing within each controller.
Attaching the CPID to the span and the logs allows us to know what changes pointed by the CPID caused the process.

The controller should be implemented to propagate CPIDs in the reconciliation loop and to output span and logs in critical processes.
This implementation requires a deep understanding of each controller, but the developer of each controller can implement it at the time of the development.

The proposed method avoids the problem of determining the completion of the change as the control plane, as discussed in Section~\ref{subsec:proposed-method-challenges}, because no process is required at the time of the completion, such as deleting the trace context on an object.
Although the cluster operator may also obtain spans and logs of unrelated changes after the original change has been completed, but they can simply filter out unrelated spans and logs by CPIDs.
In addition, although it is not the purpose of this project, by tracing the merge graph backward for the CPIDs that a particular object or span has, we can obtain several CPIDs that represent the starting point of the change. This backward tracing can be used to facilitate the investigation of how the situation of a specific span/object occurred.

\subsection{Overview}\label{subsec:proposed-method-overview}
\begin{figure}[htbp]
  \centering
  \includegraphics[keepaspectratio, width=0.8\hsize]{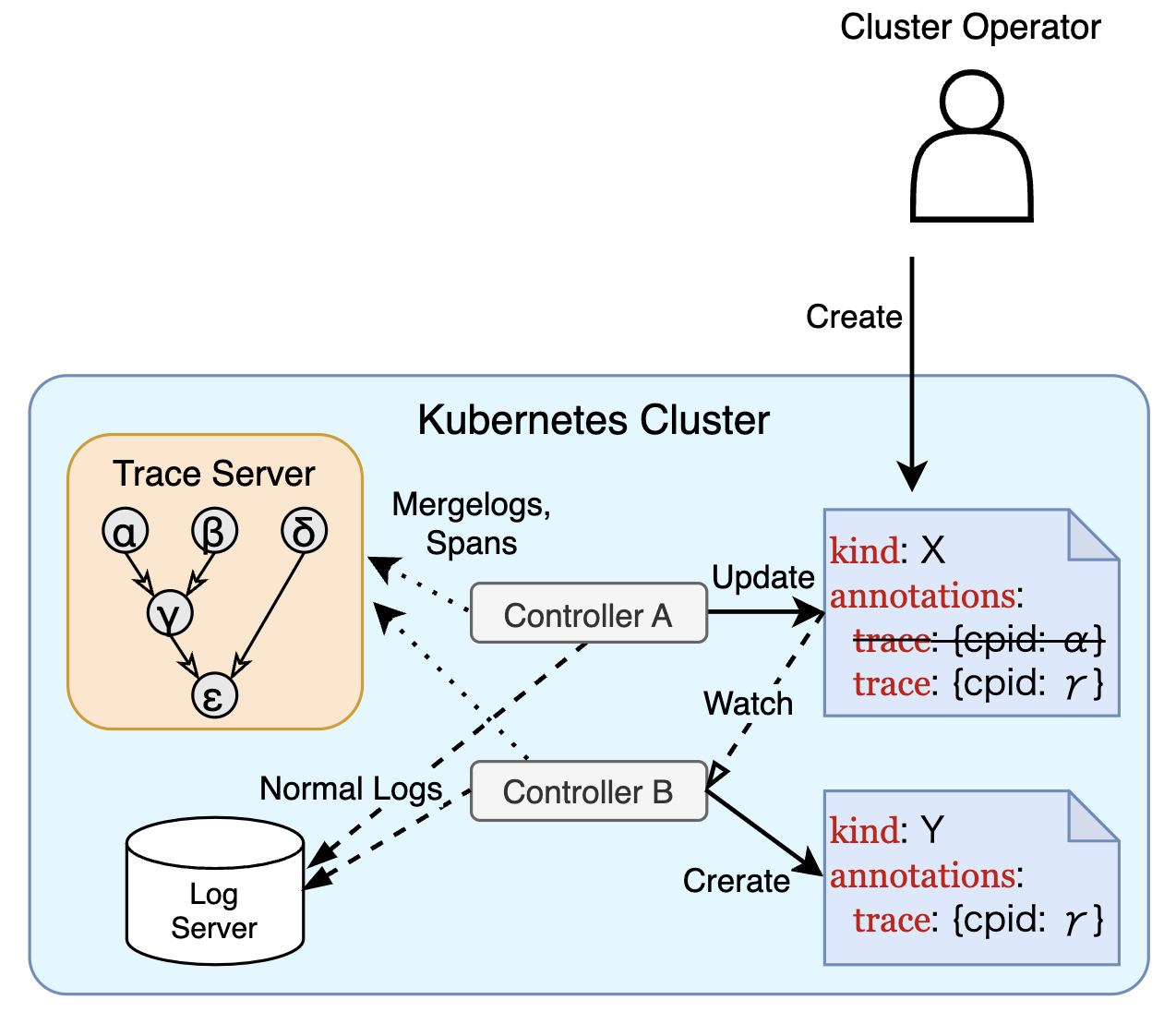}
  \caption{How the controllers work in the proposed method.
  The CPID is put on the object, and the controller sends a mergelog to the trace server at change merging, updating it with the new CPID. The controller outputs the log and span along with the CPID that caused the process.}
  \label{fig:system-overview}
\end{figure}

An overview of the proposed system is shown in Figure~\ref{fig:system-overview}.
The CPIDs are placed in the annotation field on the metadata of the object, and the controller sends a mergelog to the trace server to update the object with the new CPID when changes are merged.
Each controller sees the CPID when processing and outputs the CPID on the span and logs.
The trace server creates a merge graph from the received mergelogs and responds with the relevant CPIDs and spans in response to requests.

\begin{figure}[htbp]
  \centering
  \includegraphics[keepaspectratio, width=\hsize]{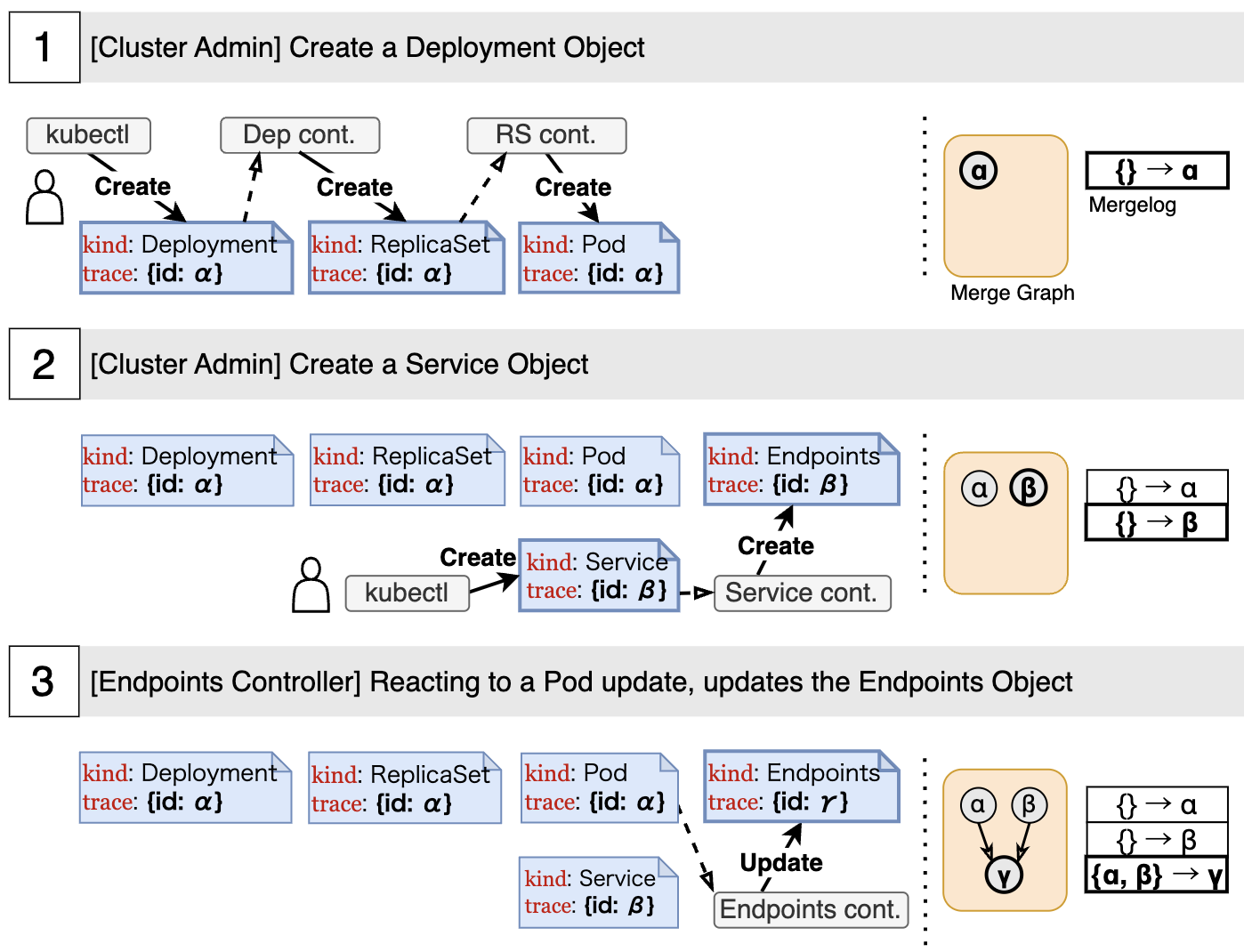}
  \caption{Proposed system propagating CPIDs and mergelog and merging graphs at each moment}
  \label{fig:propagation-with-proposed-method}
\end{figure}

Figure~\ref{fig:propagation-with-proposed-method} shows how trace context including CPID is propagated through the objects in this system.
The right side of the dotted line represents the mergelog sent to the trace server and the merge graph.

\begin{enumerate}
  \item The cluster operator creates the Deployment object. \\At this time, kubectl, a command line tool for operators, creates CPID $\alpha$. There is no change to merge it for creation, so it is placed in the Deployment object as it is. When creating a Pod object, CPID $\alpha$ is inherited as it is.
  Send a mergelog to the trace server indicating the new CPID $\alpha$ has been assigned, and a merge graph consisting only of $\alpha$ is constructed.
  \item The cluster operator creates a Service object. \\At this time, kubectl creates the CPID $\beta$, but since there is no other changes at this point, it is placed directly on the Service object. Similarly, when the Service controller creates the Endpoints object, it inherits the CPID $\beta$.
  A mergelog is sent to the trace server indicating that a new CPID $\beta$ has been assigned, and the new $\beta$ is added to the merge graph.
  \item The Endpoint controller observes the state of the Pod objects. \\In this process, it modifies the corresponding Endpoints object based on the state of the Pod objects, where the changes in $\alpha$ and $\beta$ are reflected. The Endpoints controller assigns a new CPID $\gamma$ and sends a mergelog $\{\alpha, \beta\} \rightarrow \gamma$ to the trace server, linking the original CPIDs $\alpha$ and $\beta$ with the new $\gamma$. The trace server then adds this information to the merge graph.
\end{enumerate}

\begin{figure}[htbp]
  \centering
  \includegraphics[keepaspectratio, width=\hsize]{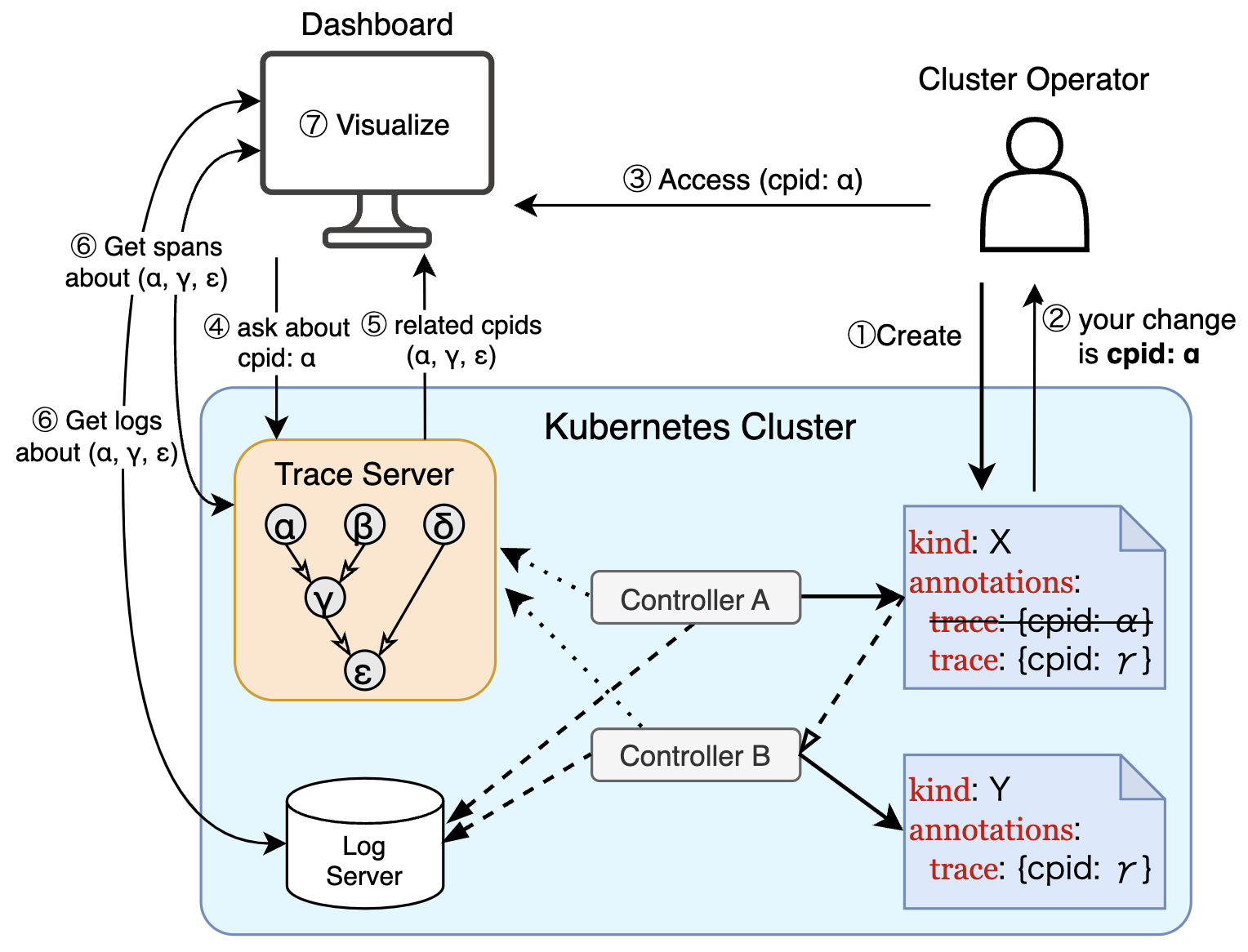}
  \caption{The process of observing change propagation with the proposed method}
  \label{fig:system-usage}
\end{figure}

Figure~\ref{fig:system-usage} shows how the operator confirms the cascading changes due to the specific update.
When the operator creates or updates an object (\ctext{1}), the CPID: $\alpha$ of the change is returned to the operator (\ctext{2}).
The operator accesses the dedicated dashboard with CPID: $\alpha$ as the key (\ctext{3}).
The dashboard queries the trace server for a list of related CPIDs derived from CPID: $\alpha$ (\ctext{4}), obtaining the list ($\alpha, \gamma, \varepsilon$) in the case of the mergelog (\ctext{5}).
The dashboard obtains the related spans from the trace server and the related logs from the log collection infrastructure based on the list of related CPIDs obtained (\ctext{6}).
The dashboard displays the obtained span and log information (\ctext{7}).

\subsection{Components}\label{subsec:components}
\subsubsection{Change Propagation ID (CPID)}\label{subsubsec:cpid}
A Change Propagation ID (CPID) is an identifier for continuously tracing changes.
In our method, only one CPID is carried per object, which is the entity for propagating changes. When each controller in the Kubernetes control plane makes a new change to an object, it assigns a new CPID and associates it with the other CPIDs of the objects that caused the change when the other CPIDs have different IDs.
The association is sent to the trace server for analysis by the operators.
To eliminate performance bottlenecks and single points of failure by assigning CPIDs, identifiers such as UUID~\cite{rfc4122} that do not conflict even if they are generated independently in a distributed environment are used to enable unique CPID generation among distributed controllers.

\subsubsection{Mergelogs and Merge Graph}\label{subsubsec:mergelog}
As described in Section~\ref{subsubsec:cpid}, only one CPID is placed on an object to keep the size of the trace context constant.
However, in the Kubernetes control plane, changes made by multiple entities are merged while they are propagated among objects, and a single change stem from multiple changes on different objects.
To fill this gap, we introduce a mergelog.
A mergelog is information that links the CPID of the original change to the newly assigned CPID placed on the object.
The mergelog is created by each controller and sent asynchronously to the trace server.
The trace server uses the mergelog to construct a merge graph that represents the ancestor-descendant relationship of CPIDs and searches for related CPIDs (CPIDs that inherit the context of that CPID) in response to requests from cluster operators.

The fields in the mergelog are listed in Table~\ref{tab:mergelog-fields}.
New CPID is a CPID that is newly generated and inherits the context of the source CPIDs, while the Source CPIDs are a list of CPIDs of the objects that the controller sees to update the object where the new CPID is assigned.
The Timestamp field is used to remove CPIDs from the merge graph.

\begin{table}[htbp]
  \caption{Fields of a mergelog}
  \small
  \centering
  \label{tab:mergelog-fields}
  \begin{tabular}{ll}
    \hline
    New CPID & newly generated CPID that inherits source CPIDs\\
    Source CPIDs & CPIDs of merge source\\
    Timestamp & Time of mergelog creation\\
    \hline
  \end{tabular}
\end{table}

Next, we will explain how to add to the merge graph using the newly received mergelog and how to remove CPIDs from the merge graph.
In the case of an addition, we create a directed edge from the vertex of each CPID in the Source CPIDs of the input mergelog to the New CPID.

Also, since the mergelog continues to grow while the trace server is running, removing CPIDs (mergelogs) from the trace server is necessary.
In doing so, it is important to ensure that the merge graph will not be destroyed and prevent other cascading changes from being traced.

The steps to remove a CPID from the merge graph are as follows.
If another CPID edges the CPID, i.e., if the information on change merging from another CPID exists, the CPID is not deleted. Otherwise, it deletes its CPID and the directed edges leaving the CPID.
The CPID's node whose input degree is newly set to 0 by the deletion of the edge is also deleted because such CPID is generated at the time of merging and is not referred to by the operator. The same process is repeated until the size of the merge graph becomes acceptable.
There are several possible criteria for deleting CPIDs when multiple CPIDs are candidates for the deletion, including the timestamp in the mergelog, the addition of other optional fields in the mergelog (e.g., a field to record which controller issued the CPID).

To trace a change, we need all of the CPIDs that represent the start of that change and the list of CPIDs generated by merging those CPIDs (related CPIDs).
To obtain the relevant CPIDs for a given CPID from a merge graph, since the merge graph is a directed acyclic graph (DAG) \footnote{If it were cyclic, the CPID generated as a new CPID would match the CPID already existing in the merge graph, but since CPIDs are generated randomly, they do not match existing CPIDs. Therefore, it is acyclic.}, we can follow the edges from the vertex representing the given CPID and enumerate the CPIDs of all the vertices that can be reached.

\begin{figure}[htbp]
  \centering
  \includegraphics[keepaspectratio, width=0.5\hsize]{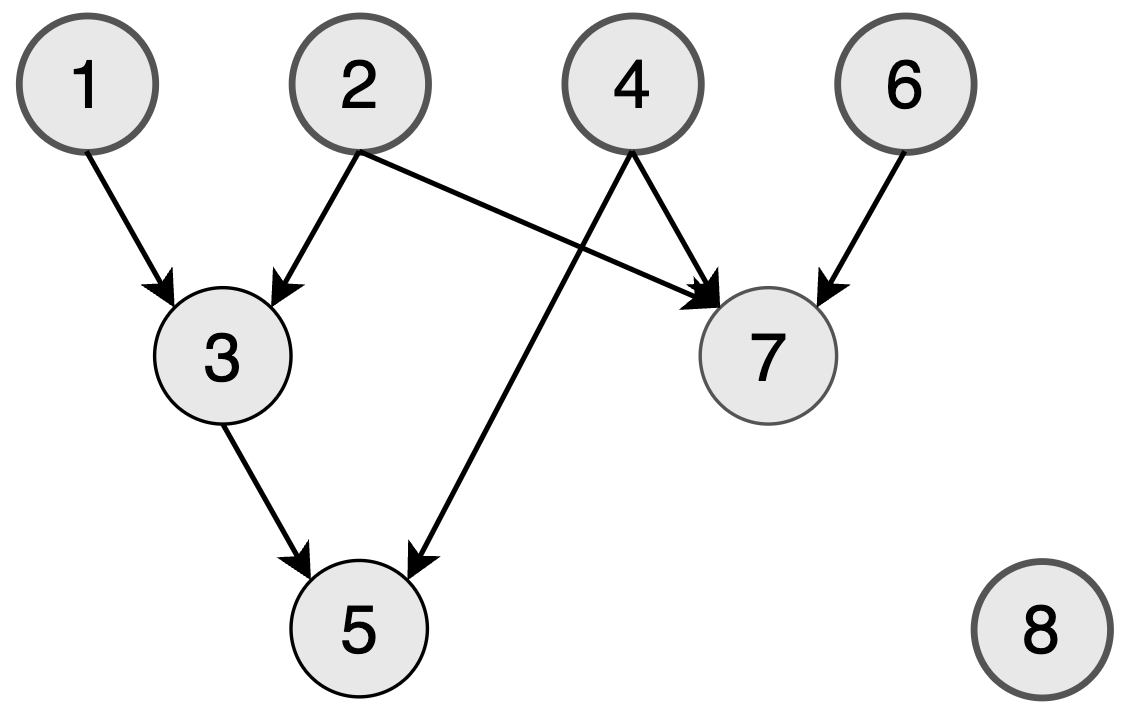}
  \caption{Example of merge graph}
  \label{fig:merge-graph-example}
\end{figure}

In the example in Figure~\ref{fig:merge-graph-example}, the input CPIDs and their associated CPIDs are shown in Table~\ref{tab:related-cpid-example}.

\begin{table}[htbp]
  \caption{Related CPID list for each input CPID of Figure~\ref{fig:merge-graph-example}}
  \centering
  \small
  \label{tab:related-cpid-example}
  \begin{tabular}{c|c}
    \hline
    Input CPID & Output related CPIDs \\ \hline
    1 & [1,3,5] \\
    2 & [2,3,5,7] \\
    3 & [3,5] \\
    4 & [4,5,7] \\
    5 & [5] \\
    6 & [6,7] \\
    7 & [7] \\
    8 & [8] \\
    \hline
  \end{tabular}
\end{table}

Here, it is possible to achieve the objectives of ``only one CPID to be placed on the object'' and ``tracing each change even when changes merge'' by selecting and replacing one of the original CPIDs instead of assigning a new CPID each time by merging. However, we generated a new CPID each time a change merges. This point is discussed in \ref{sec:discussion}.

\subsubsection{Spans}\label{subsubsec:span}

Span is a component also present in normal distributed tracing and represents a single processing step in a controller.
It is responsible for recording the start and end times of the process, as well as events during the process.
In distributed tracing in the Kubernetes control plane, which operates by observation, sequential processing is performed in each controller, and spans help identify bottlenecks in such processing.
Therefore, this proposal also employs span to trace change propagation in a distributed environment.

The fields of a span in this method are shown in Table~\ref{tab:span-fields}.
There is a CPID field to reference which change the span is linked to, and the process's start and end times are also recorded.
There is also a Service field indicating which controller the span is and the span's identifier, the Span ID.
The Parent ID slightly differs from the parent span in normal distributed tracing.
In normal distributed tracing, the trace ID and the span ID are put on the HTTP header when propagating among microservices, and the receiving microservice extracts the trace ID and span ID from the header and uses them as the trace ID and parent ID of its span.
The first span created in each controller is regarded as the root span (a span for which no parent span is specified).
When an object change is observed, the span starts with the object's CPID, and the Parent ID field is left empty.
When a child span is created from a parent span in the same controller, the child's identifier is generated as in normal distributed tracing, and the parent's Span ID is stored in the child's Parent ID field.

\begin{table}[htbp]
  \caption{Fields of a mergelog}
  \centering
  \small
  \label{tab:span-fields}
  \begin{tabular}{ll}
    \hline
    CPID & CPID of the change that caused the process \\
    Start Time & The timestamp when the process started \\
    End Time & The timestamp when the process ended \\
    Service &  The name of the controller where the process occurred \\
    Span ID & ID of this span \\
    Parent ID & ID of the parent span \\
    \hline
  \end{tabular}
\end{table}

\subsubsection{Logs by Controllers}\label{subsubsec:ordinary-logs}
The controller observes changes to objects in the reconciliation loop and performs new processing in response to those observed changes. These processes are sometimes output as logs for debugging purposes, and the logs can be used effectively to help trace the changes and trace the processes in the control plane.
However, these regular logs do not contain the necessary information to trace changes. If they were left as they were, the measurer would have to search for the necessary logs from the many logs output by many controllers. Also, since changes propagate by observation rather than invocation, it is unclear which controller's logs to check without specific knowledge.
Therefore, this method embeds CPIDs in the regular logs issued by controllers so that measurers can use the CPIDs as clues to obtain logs related to the change from the controllers involved in the change propagation without prior knowledge.

\subsubsection{Trace Server}\label{subsubsec:trace-server}

The trace server handles mergelogs and spans sent by controllers.
It communicates with the cluster operator, controllers, and dashboards for tracing through predefined interfaces.
The following are the required processes of the trace server.

\begin{itemize}
  \item Receives mergelogs as input, stores them, and constructs a merge graph
  \item Outputs all saved mergelogs
  \item Receives CPID as input and outputs the mergelogs with the related CPIDs from the information in the merge graph
  \item Receives span as input and stores it
  \item Outputs all saved spans
  \item Receives CPID as input and outputs the spans that refer to the related CPIDs from the information in the merge graph
\end{itemize}

\subsection{Ancestor CPIDs}\label{subsec:ancestor-cpid}

\begin{figure}[htbp]
  \centering
  \includegraphics[keepaspectratio, width=\hsize]{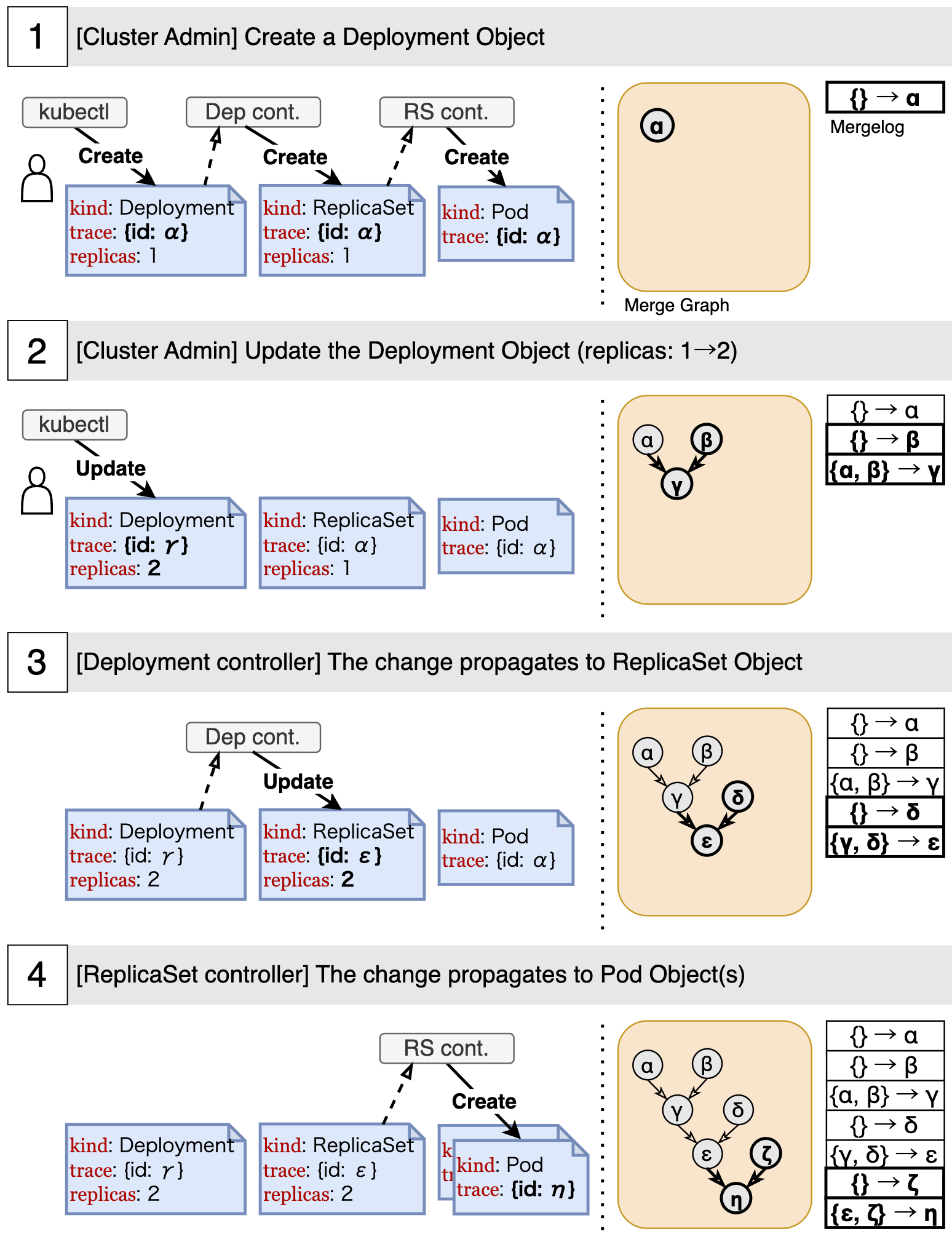}
  \caption{Situation that unnecessary merges occur many times. Each Gerrk letter represents a CPID. Since $\gamma$ already contains the information of $\alpha$, it is sufficient to replace $\alpha$ with $\gamma$, but since each controller does not know that $\gamma$ carries the information of $\alpha$, unnecessary merge processings happen.}
  \label{fig:repeated-merges}
\end{figure}

Here, we introduce a device to improve the efficiency of the merging process.
In the primary method, if an object is changed and then the same object is changed again when other changes have not merged, as shown in Figure~\ref{fig:repeated-merges}, there is a problem that many unnecessary merging processes occur in the object where the changes are propagated, which are not necessary for change tracing.
The following are some of the problems associated with many merges.

\begin{itemize}
  \item Increased processing time for change propagation as it is a critical path in  controllers
  \item Increased mergelogs sent to the trace server, which increases bandwidth usage in the cluster
  \item Increased storage size required to construct merge graphs in the trace server
  \item Larger merge graphs require more time to enumerate related CPIDs
\end{itemize}

\begin{figure}[htbp]
  \centering
  \includegraphics[keepaspectratio, width=0.4\hsize]{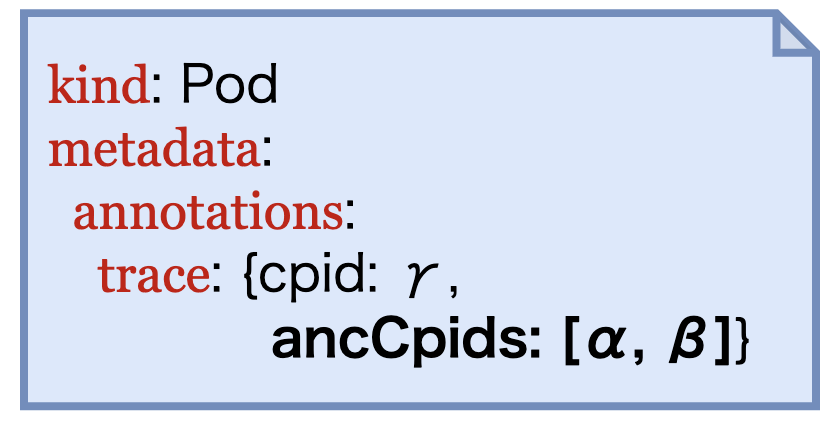}
  \caption{Example of an object with ancestor CPIDs: $\alpha$ and $\beta$ are listed as ancestor CPIDs for CPID $\gamma$.}
  \label{fig:ancestor-cpids}
\end{figure}

To eliminate such inefficiencies, we will place a constant number of ancestor CPIDs on the annotation of each object as auxiliary information, as shown in the new Figure~\ref{fig:ancestor-cpids}.
Suppose a CPID is a common descendant of all other CPIDs in the merging CPID (a CPID related to all other CPIDs). In that case, we can simply replace it with that CPID to trace the change merging, eliminating the need to generate CPIDs and send mergelogs.

Since the ancestor CPIDs are auxiliary information, they do not necessarily need to contain information on all of the N most recent ancestor CPIDs and may only contain a portion of them.
In addition, each controller may change the ancestor CPIDs unless the relationship between the CPIDs and the ancestor CPIDs is broken.
However, these logics are implemented in the instrumentation library so that it is automatically rewritten to be more efficient. The controllers usually do not need to know the ancestor CPIDs or their rewriting.
In addition, the ancestor CPIDs to be placed on objects should be the closest ancestor CPIDs on the top side of the list.
When a merge occurs and new CPIDs are generated, the ancestor CPIDs of the originators' CPIDs are added from the top of the list so that the CPIDs of the closer ancestors are placed on the list.
This idea allows us to take advantage of temporal locality, which helps to increase the percentage of merge process avoidance and handle change merging more efficiently.

The rationale for the design of putting a list of ancestor CPIDs on the object is as follows.
To determine whether any CPID can replace the CPIDs obtained as a merging source, we can examine whether all pairs of CPIDs are in an ancestor-descendant relationship.
One can check whether one CPID encompasses all other CPIDs by reviewing all previous mergelogs.
However, from the standpoint of scale and performance, it is not realistic to obtain mergelogs generated by other controllers for this purpose and to create a local merge graph in each controller to check whether merge processing is required.
Another possible method is to ask the trace server whether the set of target CPIDs can be replaced by any one of them each time a merge occurs, but the change merging process is a process on the critical path of the object change process, and communication with the trace server would cause a considerable delay.

Therefore, we chose to add CPIDs to the objects as additional information as a method that can be determined locally without additional communication.
Since the list of ancestor CPIDs does not include all ancestor CPIDs, it is not always possible to find cases where merging is not necessary, but it is effective in simple cases such as Figure~\ref{fig:repeated-merges}, where the decision can be made based on the most recent merge only.
It may seem meaningless to add additional CPID information to a single object with only one CPID to prevent the size of the trace context from becoming too large, but since the number of ancestor CPIDs is fixed to $N$ before the cluster is created, the size of the trace context to be included in the object will not increase without limit.
The improvement in the number of mergelogs by introducing ancestor CPIDs will be evaluated in \ref{sec:evaluation}.

\section{Implementation}\label{sec:implementation}
We implemented part of the proposed system to verify that the proposed method works in practice.

\subsection{Environment}\label{subsec:implementation-env}
We use Go 1.20 and kind 0.20.0 for implementation and measurement, and Protocol Buffers~\cite{protobuf3} and gRPC~\cite{grpc} for defining data format and interfaces between components.
The Kubernetes cluster is created using kind~\cite{kind} on a virtual machine with the specifications shown in Table~\ref{tab:machine-spec}.
The controller implementation is based on the August 2023 commit \footnote{https://github.com/kubernetes/kubernetes/commit/02e51b27a9a40bd10094d4d87d90aff78ace171e} of the Kubernetes source code.

\begin{table}[htbp]
  \centering
  \small
  \caption{Specification of host and virtual machines}
  \label{tab:machine-spec}
  \begin{tabular}{lll}
    \hline \hline
                     & host machine   & virtual machine    \\
    \hline
    OS               & Ubuntu20.04.3  & Ubuntu20.04.6 \\
    Kernel           & 5.4.0          & 5.4.0         \\
    CPU(vCPU)        & AMD EPYC 7402P & 8             \\
    Memory           & 196GiB         & 24GiB         \\
    \hline \hline
  \end{tabular}
\end{table}

\subsection{Instrumentation Library}\label{subsec:instrumentation-library}

Processes such as establishing communication with the trace server, generating CPIDs, sending mergelogs and spans, and creating child spans are common to each controller.
The instrumentation library provides these operations necessary for instrumentation.

In Section~\ref{sec:proposed-method}, we stated that CPIDs must not conflict even if distributed components generate them independently. Therefore, this implementation uses UUID version 4~\cite{rfc4122} as the CPID.
UUID version 4 is also used as a Span ID.

The instrumentation library provides the following functions.
The trace context here refers to a structure consisting of a CPID and a list of ancestor CPIDs.

\begin{itemize}
  \item Inject trace contexts on a variable
  \item Extract trace contexts from a variable
  \item Inject a trace context on an object
  \item Extract a trace context from an object
  \item Generate a root CPID and it to the trace server
  \item Merge multiple trace contexts to generate a new CPID and send a mergelog if necessary
  \item Start and end a span
\end{itemize}

Algorithm~\ref{alg:build-cpid-graph} is an algorithm that, given multiple trace contexts (\texttt{tctxs}), constructs an ancestor relationship graph of CPIDs based on their CPIDs and ancestor CPIDs to determine if merging is necessary.
If the graph returned by this algorithm has only one root, the CPID of that root can be determined to be the related CPID of all other CPIDs, and merge processing is unnecessary. If the graph has more than one root, it is necessary to merge those CPIDs to generate a new CPID.

\begin{algorithm}[htbp]
  \footnotesize
  \caption{Build CPID Graph}
  \label{alg:build-cpid-graph}

  \begin{algorithmic}[1]
  \Require{tctxs: list of trace context}
  \Ensure{cpidGraph: graph of CPID (dictionary)}

  \Function {BuildCpidGraph}{tctxs}
    \State cpidGraph $\gets$ \{\}    // create an empty cpidGraph
    \State // for each tctx in tctxs, add it to cpidGraph
    \ForAll {tctx in tctxs}
      \State cpidGraph $\gets$ addCPID(cpidGraph, tctx)
    \EndFor
    \State \Return cpidGraph
  \EndFunction
  \State
  \Function {addCPID}{cpidGraph, tctxs}
    \State // Step 1: check if any of tctx.AncCpids matches to any roots.
    \State // If so, add the ancestors of the root to cpidGraph[tctx.cpid] 
    \State // and delete the root from cpidGraph.roots. If not, do nothing.
    \ForAll {key in cpidGraph.keys}
      \If {key is in tctx.AncCpids}
        \ForAll {val in cpidGraph[key]}
          \If {key is not in tctx.AncCpids}
            \State // append because the end side is older ancestor
            \State add val to tctx.AncCpids
          \EndIf
        \EndFor
        \State delete cpidGraph[key]
      \EndIf
    \EndFor
    \State // Step 2: check if tctx.Cpid is included in the roots of cpidGraph.
    \State // If so, add tctx.AncCpids as values of the root. If not, add tctx to cpidGraph's root.
    \State cpidIsIncluded $\gets$ false
    \ForAll {key in cpidGraph.keys}
      \If {key $=$ tctx.Cpid $\parallel$ tctx.Cpid is in cpidGraph[key]}
        \For {$i = len(tctx.AncCpids) - 1$ to $0$}
          \If {tctx.AncCpids[i] is not in cpidGraph[key]}
            \State // prepend because the start side is newer ancestor
            \State add tctx.AncCpids[i] to front of cpidGraph[key] 
          \EndIf
        \EndFor
        \State cpidIsIncluded $\gets$ true
      \EndIf
    \EndFor
    \If {not cpidIsIncluded}
      \State cpidGraph[tctx.Cpid] $\gets$ tctx.AncCpids
    \EndIf
    \State \Return cpidGraph
  \EndFunction
  \end{algorithmic}
\end{algorithm}

\subsection{Instrumentation of Controllers}\label{subsec:instrumentation-of-controllers}
Some controllers are instrumented using the instrumentation library in Section~\ref{subsec:instrumentation-library}.

\noindent
\textbf{kubectl}\quad 
kubectl is a command line tool that communicates with the API server to operate a Kubernetes cluster and, given an object definition, sends it to the API server. This time, we have implemented to generate a CPID, send a mergelog, and create an object with that CPID on it when the \texttt{--trace} flag is set. Also, since this is the starting point of object creation and modification, the root CPID generated by kubectl is notified to the command executor (cluster operator) so that the change propagation regarding the change can be traced.

\noindent
\textbf{Deployment Controller and ReplicaSet Controller}\quad 
When an object is changed, each controller merges the CPIDs of the objects observed in that reconciliation loop and sends a mergelog. The trace context obtained from the merge processing is placed on the object to be changed.
In addition, the span is instrumented to capture the processing time for functions that are expected to take a long time.

\noindent
\textbf{Scheduler}\quad 
The Scheduler was instrumented to generate spans by the Pod's CPID when scheduling Pod objects.

\subsection{Trace Server}\label{subsec:impl-trace-server}

The trace server is implemented in Go mainly because the Kubernetes library is fully maintained.
The protocol buffer compiler generates the Go language data type and communication interface code from the aforementioned scheme definitions, and the trace server is implemented using this code.
In addition, the trace server is deployed as a Kubernetes Deployment object to take advantage of Kubernetes' fault tolerance and name resolution features.
The structure that stores the span, mergelog list, and merge graph is implemented with exclusive control to prevent conflicts by concurrent accesses.

\subsection{Dashboard}\label{subsec:dashboard}
\begin{figure}[htbp]
  \centering
  \includegraphics[keepaspectratio, width=\hsize]{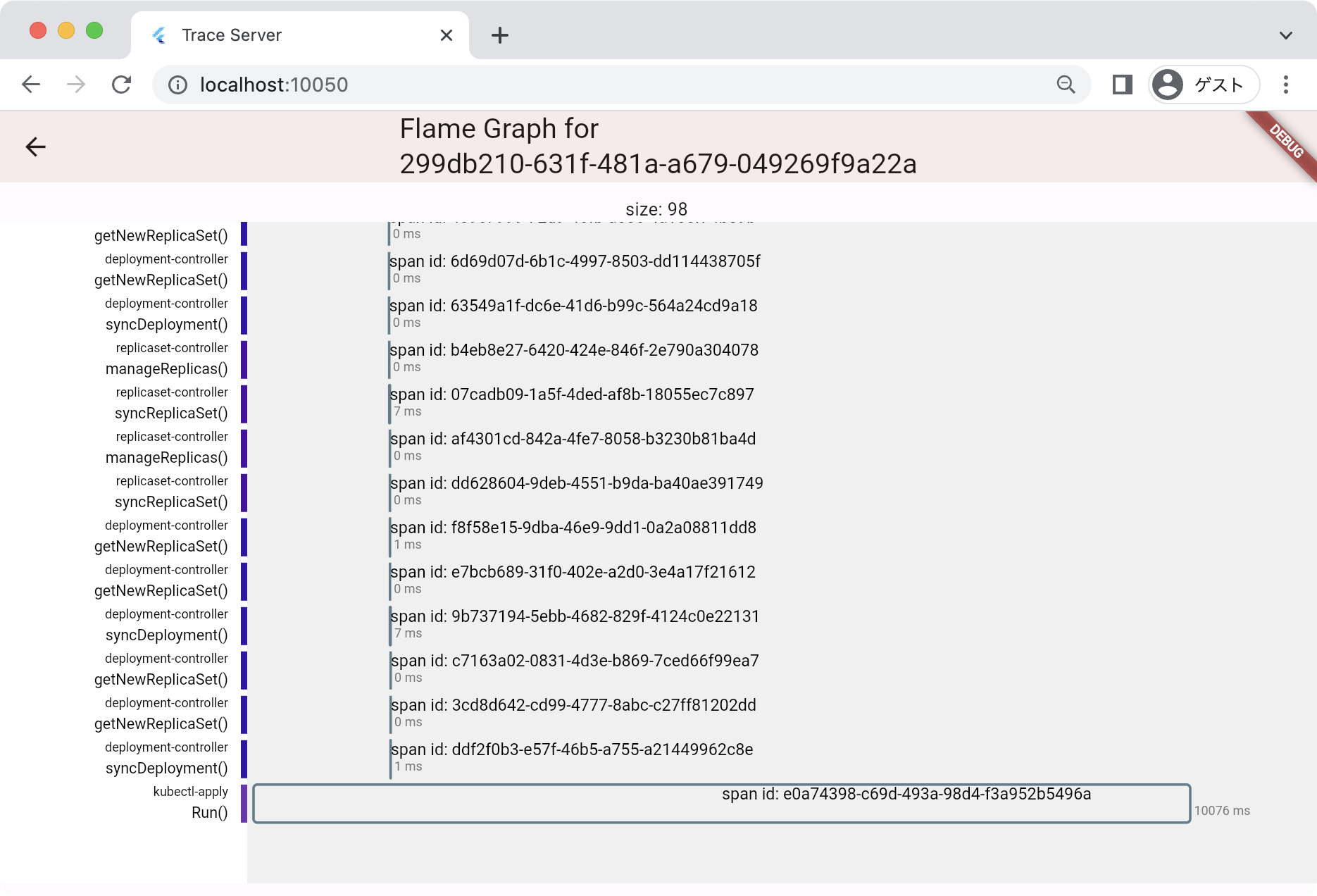}
  \caption{Given a CPID, the dashboard displays the spans associated with that CPID in a flame graph format. Spans from kubectl, Deployment controller, ReplicaSet controller, and Scheduler are shown on this page (a part of spans is shown in the figure).}
  \label{fig:dashboard-flame-graph}
\end{figure}

A dashboard intended to be used by measurers for investigations is also implemented.
The dashboard was implemented using Flutter and runs on a web browser.
On the page shown in Figure~\ref{fig:dashboard-flame-graph}, when a CPID is input, the span associated with that CPID is displayed in a frame graph.
Such flame graphs allow users to see which controllers and processes take a long time.
We can see spans from all the instrumented controllers; kubectl, Deployment controller, ReplicaSet controller, and Scheduler.
It means that the CPIDs are propagated through these controllers properly.

\section{Evaluation}\label{sec:evaluation}

\subsection{Evaluation with Measurement}\label{subsec:evaluation-by-measurement}

\subsubsection{Overhead}\label{subsubsec:system-overhead}
We evaluate how much overhead this system incurs by measuring the time required for change propagation and the CPU and memory usage of the trace server.
The pre-implementation code is used here, as described earlier in Section~\ref{subsec:implementation-env}.

\noindent
\textbf{Scenario of changes}\quad
The object change scenario for this measurement is as follows.
The measurements were taken ten times in the two states: before (original) and after (instrumented) instrumentation.
In each case, the time required to run this scenario was measured, and for the instrumented case, the CPU and memory usage of the trace server was also measured.

\begin{enumerate}
  \item Create a Deployment with 10 replicas of Pods and wait for all Pods to be ready.
  \item Change the number of Pod replicas to 20 and wait for all Pods to be ready.
  \item Change the number of Pod replicas to 30 and wait for all Pods to be ready.
  \item Change the number of Pod replicas to 40 and wait for all Pods to be ready.
  \item Change the number of Pod replicas to 50 and wait for all Pods to be ready.
\end{enumerate}

In this scenario, there are five change propagations that propagate from Deployment to ReplicaSet and then to Pod.
The controllers that perform processing between these are kubectl, Deployment Controller, ReplicaSet Controller, Scheduler, and kubelet, and four controllers except kubelet have been instrumented.

\noindent
\textbf{Results}\quad
The average CPU and memory usage of the trace server during the scenario was 18 milliCPU  \footnote{https://kubernetes.io/docs/tasks/configure-pod-container/assign-cpu-resource} for CPU and 29 Mi bytes for memory, respectively.
For comparison, the API server's average CPU and memory usage during the no-object-changes period was 68 milliCPU and 45 Mi bytes of memory, respectively.
As can be seen by the comparison with the API server, the trace server's CPU and memory usage is insignificant and has little impact on the other workloads.

The average time required to run the scenario was 32.09 seconds before instrumentation and 73.84 seconds after instrumentation.
Since five change propagations were executed, the time per propagation was 6.42 seconds before and 14.77 seconds after instrumentation.
Although this time may vary depending on how much instrumentation is used (e.g., number of spanning points, number of controllers instrumented in the change propagation path), in this implementation, the time required for change propagation increased by about 8 seconds due to the instrumentation.
However, considering that this method will enable debugging of delays of several minutes caused by controller bugs, increasing the time by several seconds in regular cluster operation is considered acceptable.
This overhead is also expected to be reduced by optimizing the instrumentation library and instrumentation.

\subsubsection{Change in the Number of Mergelogs due to a Change in the Upper Limit $N$ of Ancestor CPIDs}\label{subsubsec:mergelog-number-by-ancCpid-limit}
To evaluate the effect of ancestor CPIDs on mergelog reduction, we check how changing the maximum number of ancestor CPIDs $N$ placed on an object affects the number of output mergelogs.

\noindent
\textbf{Scenario of changes}\quad
The object change scenario for this measurement is as follows.
Measurements were taken ten times each for each $N=0$, $1$, $2$, $3$, $5$, $10$, $15$, $20$, $30$.
Note that there is a fixed wait time between each step for the change to propagate to the pod at each step.

\begin{enumerate}
  \item The following processes with five Deployment objects are repeated four times.
  \begin{enumerate}
    \item Change the number of Pod replicas of all Deployments to 1 (total number of pods is 5)
    \item Change the number of Pod replicas of all Deployments to 3 (total number of pods is 15)
  \end{enumerate}
\end{enumerate}

In this scenario, the Deployment object is updated seven times, excluding the initial creation, and for each Deployment update, the Deployment object before the update and the root CPID from the update are merged.
Since each step updates the CPID of the Deployment object, the Deployment controller that observes the change performs a merge processing of the CPID of the new and old Deployment object of the corresponding ReplicaSet object.
Suppose the ReplicaSet object's CPID is included in the Deployment object's ancestor CPIDs. In that case, merge processing is unnecessary, and the ReplicaSet object's CPID is simply replaced with the Deployment object's CPID. If not, the Deployment controller cannot determine that it is an ancestor, so it generates a new CPID and sends a mergelog.

\begin{figure}[htbp]
  \centering
  \includegraphics[keepaspectratio, width=0.8\hsize]{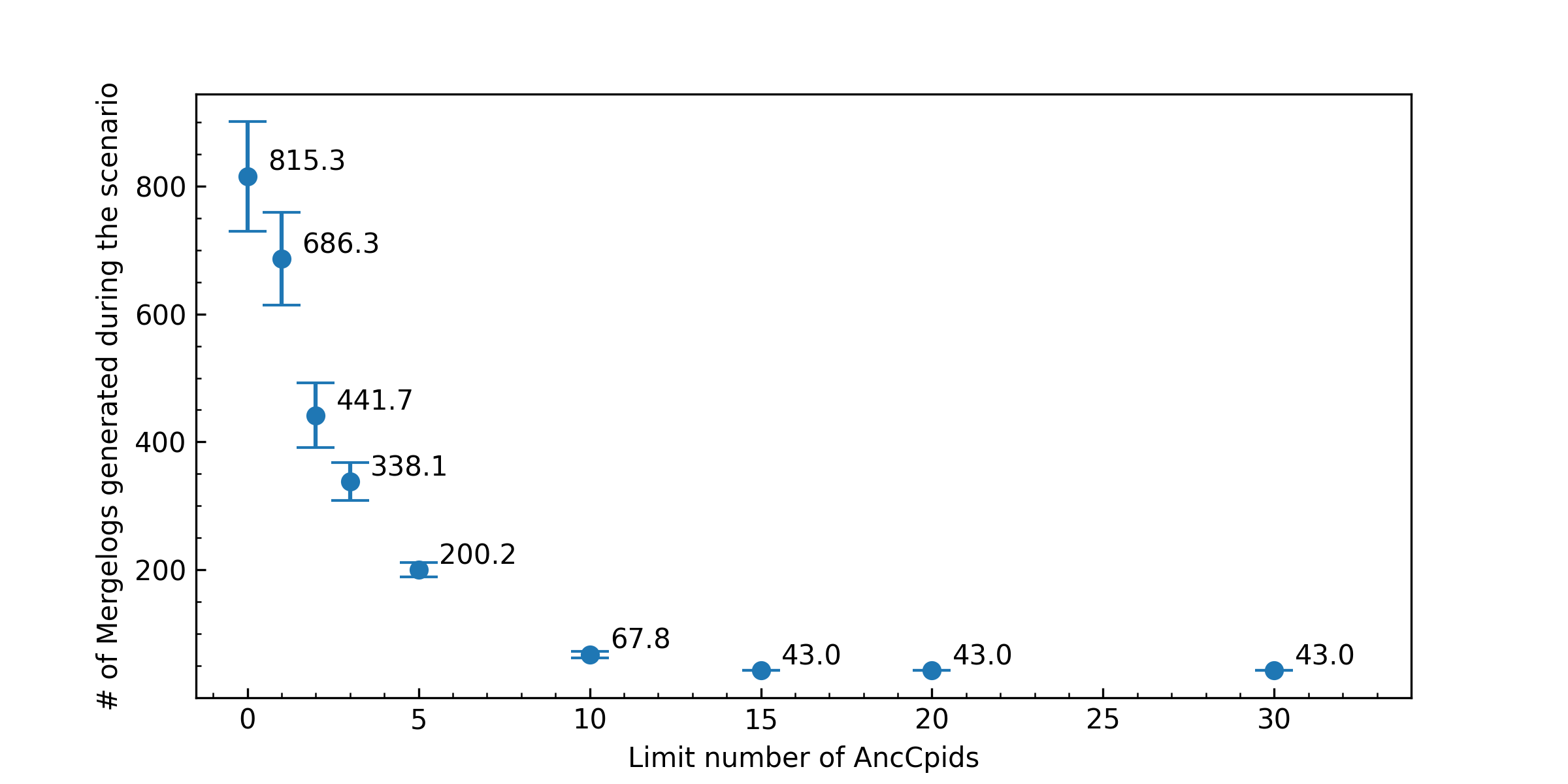}
  \caption{The number of output mergelogs as the limit number of ancCpid ($N$) varies. }
  \label{fig:mergelog-number-by-ancCpid-limit}
\end{figure}

\noindent
\textbf{Results}\quad
Figure~\ref{fig:mergelog-number-by-ancCpid-limit} shows the number of mergelogs sent.
The horizontal axis is the upper limit of $N$, and the vertical axis is the number of mergelogs sent to the trace server during the scenario.
The data points represent the mean, and the error bars represent the standard error.

The number of mergelogs decreases as $N$ increases.
Compared to the case where ancestor CPIDs are not used ($N=0$),
the number of mergelogs is about 25 $\%$ with $N=5$ and 8 $\%$ with $N=10$.
It can then be read that the number of mergelogs reached the lower limit of the number of mergelogs after $N=15$ in this number of object changes (which is seven).
This is because all the ancestors' information can be placed as ancestor CPIDs without overflowing at the upper limit.
This result shows that ancestor CPIDs can significantly reduce the number of mergelogs, even if the number is small.

\subsection{Other Evaluations}\label{subsec:evaluation-without-measurement}

\subsubsection{Compliance with Requirements}\label{subsubsec:compliance-with-requirements}
In this section, we confirm that the proposed method satisfies the requirements described in Section~\ref{subsec:proposed-method-challenges} (excluding ``The performance degradation to the cluster is minor'').

\noindent
\textbf{Spans and logs related to a given change can be easily retrieved later.}\quad 
As confirmed in \ref{sec:implementation}, for the processes that have been implemented, it is possible to retrieve the mergelogs and spans sent to the trace server even as the changes propagate through Deployment, ReplicaSet, and Pod.
In addition to that, when checking the CPIDs referred to by the spans obtained by CPID search, spans referring to the related CPIDs from which the input CPIDs were derived can also be obtained, and it can be confirmed that change merging can be traced correctly for change propagation starting from Deployment.
In Section~\ref{subsec:capability-of-tracing-change-propagation}, we discuss what kind of change propagation can be traced by this method and what cannot.

\noindent
\textbf{There should be low impact in the presence of non-instrumented controllers.}\quad
Because objects that unimplemented controllers modify are not given new CPIDs by merging changes, the context of the changes held by other objects referenced at the time of modification is not propagated from there.
However, since the CPID originally held by the object remains unchanged, when the object is subsequently processed by another implementation controller, the merging process is performed as usual, and the trace contexts will be successfully propagated.
Thus, even a change passing through a non-instrumented controller does not destroy other trace contexts, so the impact of the presence of non-instrumented controllers is rationally small.

\section{Discussion}\label{sec:discussion}

\subsection{Extensibility of the Proposed Method}\label{subsec:extensibility-of-proposed-method}

Although this study targets Kubernetes, the proposed method can be extended to other systems.
The behavior of the proposed method can be generalized as follows.
An autonomous entity (controller in Kubernetes) reads trace contexts from something that caused the process (object in Kubernetes), embeds a trace context in the target of the change (object in Kubernetes), and outputs information (mergelog) which links those read and embedded trace contexts.

Heat~\cite{openstack-heat} is a non-Kubernetes system that enables the declarative management of OpenStack~\footnote{https://www.openstack.org/} components.
Heat reads templates created by operators.
Heat modules read the resources (machine configuration and IP address settings) described in the templates and control them using OpenStack's API.
When a change is made to the created resources or the template, it detects and reflects the change.
In this case, the modules in Heat are the entities that operate autonomously, and the templates and resources are the cause and target of the changes.
It would be possible to trace the change propagation by instrumenting the modules in Heat to output information about which template was changed and which process was performed.

\subsection{Comparison of Merging Methods}\label{subsec:cpid-merge-or-replace}
As described in Section~\ref{subsubsec:mergelog}, one possible method of updating the CPID to be placed on the object upon change merging is to select one of the CPIDs of the merging source as the CPID to be placed on the object, in addition to the method of assigning a new CPID.
In the method of using an existing CPID, a log indicating that one CPID has been replaced by another CPID (replacement log) is sent to the trace server.
The replacement log can still achieve the objectives of ``only one CPID on the object'' and ``tracing beyond merging of changes''.
In this section, we examine these two methods and discuss the reasons for adopting the former method.

\begin{figure}[htbp]
  \centering
  \includegraphics[keepaspectratio, width=\hsize]{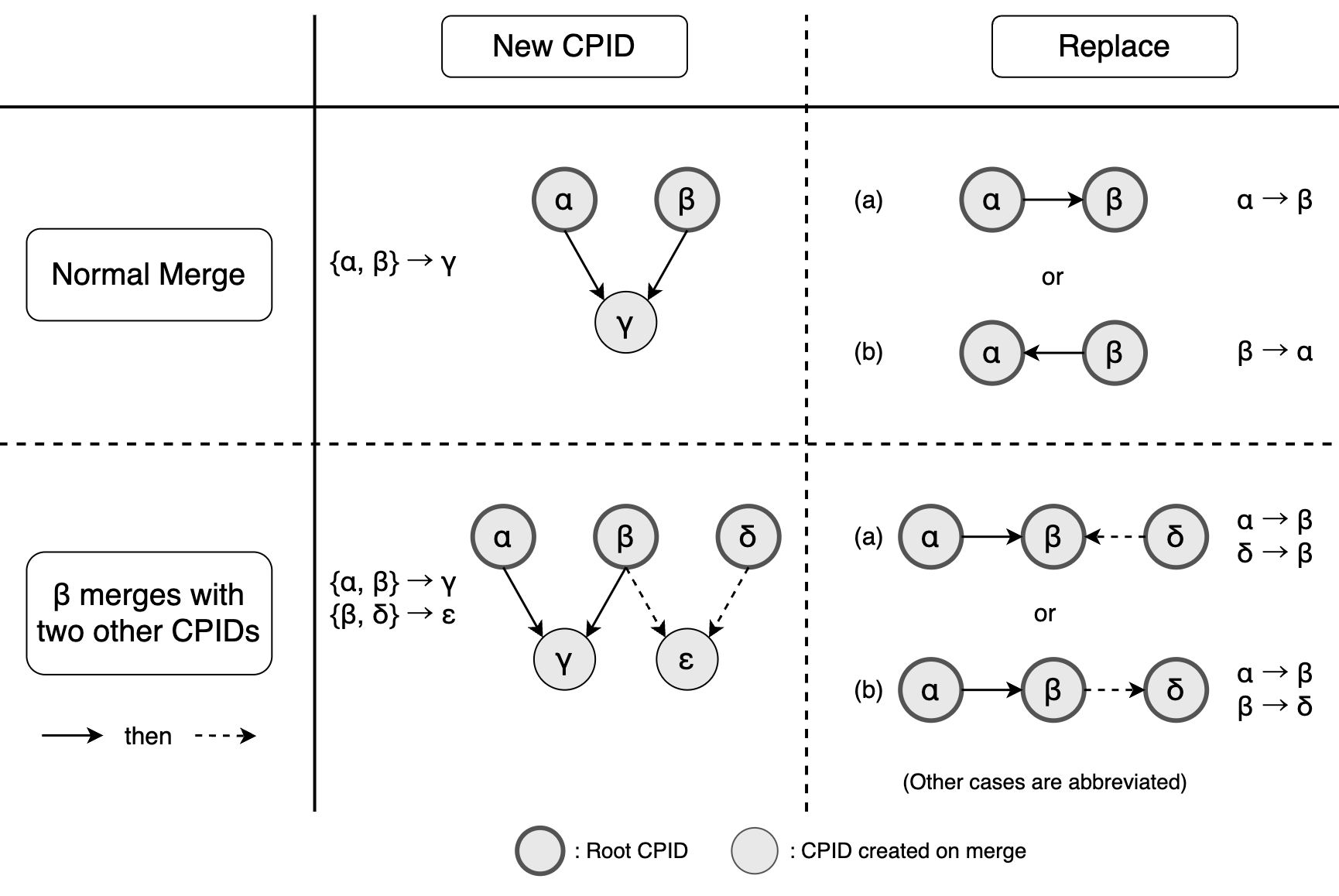}
  \caption{Comparison of two methods: new CPID and existing CPID in two cases. In the above case, $\alpha$ and $\beta$ are merged. In the below case, $\alpha$ and $\beta$ are merged followed by $\gamma$ and $\delta$.}
  \label{fig:merge-or-replace}
\end{figure}

The replacement method has a small processing time overhead because it does not generate a new CPID, and the size of the replacement log is also small.
However, there is the problem of being unable to trace changes accurately.
For example, consider the situation shown in Figure~\ref{fig:merge-or-replace}.
The above case shows where change indicating $\alpha$ and $\beta$ join together.
When the measurer searches for the span log associated with $\alpha$, if $\alpha$ is replaced by $\beta$, $\beta$ will appear as the associated CPID of $\alpha$, but the span log produced by $\beta$ before merging with $\alpha$ does nothing to do with $\alpha$.

However, it is possible to remove those related to the previous $\beta$ by searching using the timestamps of the replacement logs.
On the other hand, when generating a new CPID, $\alpha$ and $\beta$ can be distinguished, so $\beta$ is not searched as a related CPID of $\alpha$ when searching the span logs related to $\alpha$.

Next, in the lower part of the figure, after $\alpha$ and $\beta$ are merged, $\gamma$ and $\delta$ are merged.
In the case of consistently generating a new CPID on the left side, we can distinguish changes originating from $\alpha$, $\beta$, and $\gamma$, respectively.
If the existing one replaces the CPID as in (a), then when the span log with $\beta$ is observed after the two merges have occurred, even if the timestamp of the replacement log is used, it is not possible to tell whether it is related to $\alpha$ or to $\gamma$. The result is that we get a span log unrelated to $\alpha$.
As in the case of $\beta$ in (b), it is possible to design the CPID so that once it replaces another CPID, it does not replace the other CPID again, but in this case, it is not possible to deal with the case where CPIDs that have already taken in changes merge. A design in which a new CPID is generated only in this case could be considered, but this is not very easy, as it would require a flag to indicate whether a CPID has already merged with another CPID or not. 

Based on the above considerations, we adopted the method of generating a new CPID each time.

\subsection{Change Propagation Traceability}\label{subsec:capability-of-tracing-change-propagation}
This section discusses change propagation that can be traced by this method and that cannot be traced by this method.
The proposed method allows change propagation to be traced by assigning CPIDs to objects.
Since changes in the Kubernetes control plane are done by changing the fields of an object, in general, if change propagation is done by creating an object or changing the fields of an object, it is possible to trace the change propagation using the proposed method.
However, some cases are not traceable. Examples of untraceable change propagation are discussed below.

First, some cases are inherently untraceable: changes that are propagated due to the absence of an object.
Changes triggered by the deletion or non-existence of an object cannot be traced by this method if the CPID of the deleted object that caused it cannot be referenced.

Second, changes that are not represented as objects in Kubernetes cannot be traced. For example, communications and file modifications caused by an application in a container cannot be traced by this method because they are not represented as Kubernetes objects.
However, these changes occur outside of the Kubernetes control plane, and because they are not represented as objects, they will not trigger other controllers, and they are outside the scope of our method. Conversely, this method can trace their changes once they are represented as objects in Kubernetes.z

\subsection{Reliability of the Proposed Method and Effect to the Cluster}\label{subsec:reliability-and-effect-to-cluster}

\subsubsection{System-wide Impact of Trace Server Reliability}

The trace server accepts mergelogs and spans from controllers and builds the merge graph.
Therefore, if the trace server is stopped, mergelogs and spans that were received during that time will be recovered. Then, tracing or retrieving spans beyond the change merges that occurred during that time will be impossible.
When the trace server is down, each controller can avoid blocking its primary process by sending messages to the trace server asynchronously.
One idea is to introduce retransmission logic in the controller to improve the reliability of change propagation tracing.
However, since change tracing is only an auxiliary function for control plane debugging, it should not affect the performance of the original cluster systems.

\subsubsection{Macilious Controllers and Tolerance to Them}\label{subsec:malicious-controllers}
While deploying third-party resources and controllers can add new functionality to Kubernetes, some may be malicious or contain bugs.
We discuss the negative impact of this method of abuse on change tracing and its potential impact on the underlying system.

\noindent
\textbf{Incorrect modification of trace contexts (CPIDs and ancestor CPIDs) in objects}\quad 
If the trace context on an object is deleted or an irrelevant and incorrect CPID is placed, the changes propagated to that object will no longer be traceable.
Since the ancestor CPIDs are only used as an aid in merging, the removal of ancestor CPIDs does not affect the tracing of change propagation.
However, incorrect ancestor CPIDs could cause the merge to be considered unnecessary and prevent a typical controller from sending the necessary mergelogs.

\noindent
\textbf{Sending incorrect mergelogs}\quad 
Merge graphs in the trace server are updated when mergelogs are sent. If a mergelog is sent that ties unrelated CPIDs together, changes that are not related are merged, preventing effective investigation.

\noindent
\textbf{Massive amount of mergelogs and spans}\quad
Suppose many mergelogs or spans are sent, such as when many objects are changed due to a controller bug or failure. In that case, they can consume the resources of the trace server and the network bandwidth of the control plane and interfere with other legitimate communications.
In normal distributed tracing, spans sent from each service are sampled at each trace so that the amount handled can be kept below a certain level regardless of the amount of input.
In the control plane, however, sampling is undesirable because if any part of the mergelog is missing, the change propagation in that part will be interrupted. In addition, it is impossible to sample a span for every change because no unit represents a change that always ends, such as a trace ID in normal distributed tracing.

As described above, it is challenging to ignore malicious mergelogs and spans, and therefore, tracing will be difficult in an environment where malicious controllers are present.
However, since no controller determines its behavior based on an object's trace context, even if a malicious controller or a controller that contains bugs in its instrumentation exists, it will not affect the non-instrumentation parts of other controllers.

\section{Conclusion}\label{sec:conclusion}
The Kubernetes control plane, with its declarative configuration management, does not automatically trace change propagation across controllers, making it difficult to investigate the time required for change propagation and the cause of failures.
In this paper, we proposed a distributed tracing method that allows change propagation to be traced by placing only one Change Propagation ID (CPID) on an object, generating a new CPID when changes merge, and placing the CPID in the regular logs issued by the span and controller. We also proposed an idea to reduce the number of mergelogs issued by placing ancestor CPIDs on the objects as auxiliary information for merging decisions.
The controllers must be instrumented to propagate trace contexts in the proposed system.
This makes it possible to trace any change, compared to existing systems where the types of observable resources and state transitions are fixed.
It also does not require much understanding of the controller for the measurer, although the instrumentation requires some effort from its experts.
We have implemented the core functionality of the proposed method and showed that the time required for change propagation can be measured in an experimental environment.
We also confirmed that the performance impact of this system on clusters is not significant.
Furthermore, we showed that even a few ancestor CPIDs can significantly reduce the number of mergelogs.

Confirmation of the proposed method's scalability in larger clusters and measurement in more practical situations by implementation on third-party controllers are our future tasks.

\bibliographystyle{ACM-Reference-Format}
\bibliography{sigconf}

\end{document}